\documentclass[a4paper,fleqn]{cas-dc}

\usepackage[numbers]{natbib}

\def\tsc#1{\csdef{#1}{\textsc{\lowercase{#1}}\xspace}}
\tsc{WGM}
\tsc{QE}

\newcommand{\beq}{\begin{equation}}
\newcommand{\eeq}{\end{equation}}
\newcommand{\bdis}{\begin{displaymath}}
\newcommand{\edis}{\end{displaymath}}
\newcommand{\bea}{\begin{eqnarray}}
\newcommand{\eea}{\end{eqnarray}}

\ExplSyntaxOn \cs_gset:Npn \__first_footerline: { \group_begin: \small \sffamily \__short_authors: \group_end: } \ExplSyntaxOff

\begin{document}
\let\WriteBookmarks\relax
\def\floatpagepagefraction{1}
\def\textpagefraction{.001}

\shorttitle{Electric field effect on spin waves}    

\shortauthors{V. Basso, P. Ansalone, A. Di Pietro}  

\title [mode = title]{Disentangling electric field effect on spin waves in ferromagnetic insulators}  



%

\author[1]{Vittorio Basso}

\cormark[1]

\ead{v.basso@inrim.it}

\affiliation[1]{organization={Istituto Nazionale di Ricerca Metrologica},
            addressline={Strada delle Cacce 91}, 
            city={Torino},
            postcode={10135}, 
            country={Italy}}

\author[1]{Patrizio Ansalone}

\author[1,2]{Adriano {Di Pietro}}

\affiliation[2]{organization={Politecnico di Torino},
            addressline={Corso Duca degli Abruzzi 24}, 
            city={Torino},
            postcode={10129}, 
            country={Italy}}

\cortext[1]{Corresponding author}



\begin{abstract}
In this paper we extend the micromagnetic theory of magnetostatic surface waves in insulating ferromagnetic thin films to include the applied electric field effects. We start by identifying the two main effects on the dispersion relation: the first one is of relativistic nature and emerges as a consequence of the Ahronov-Casher effect, while the second one is a consequence of the electric field induced symmetry breaking operating at the level of magnetic exchange interactions. We support our theory by comparing its predictions with experimental data on ittrium iron garnet thin films taken from the literature. The main result is to evidence the limitations of using the same value of the applied electric field to address both effects and to emphasize that crystal symmetry breaking due to the applied electric field brings about the contributions of the crystal field and determines different amplitudes for the two effects.
\end{abstract}




\maketitle


\section{Introduction}

The effect of electric field on the magnetization of ferromagnets is currently an active topic of research because of its significant implications for future magnetic memory devices \cite{Rovillain-2010, Liu-2011, Zhang-2014, Amiri-2015}. It has been shown that through an electric field, it is possible to tune the phase of the magnetostatic spin waves \cite{Zhang-2014} or to induce an additional, voltage-induced, anisotropy in magnetic memory elements \cite{Amiri-2015}. The electric-field control of spin waves has been exploited in multiferroics \cite{Rovillain-2010} to design novel magnonics architectures and in insulating ferromagnets as a method to perform logic operations \cite{Liu-2011}. The microscopic origin of the electric field effect on ferromagnets has been discussed by several authors. Cao et al. \cite{Cao-1997} have evidenced the role of an electric field in the acquisition of an additional phase of the spin waves in analogy with the Aharonov-Casher effect \cite{Aharonov-1984}, while Mills and Dzyaloshinskii \cite{Mills-2008} have demonstrated that, through the flexoelectric interaction, the electric field has an effect similar to the Dzyaloshinskii Moryia interaction (DMI) leading to an extra term in the dispersion relation which is linear in the wavenumber \cite{Moon-2013}. Indeed, experiments show that the phase acquired by the spin wave has two components: one is frequency independent like in the Aharonov-Casher effect and the other is approximately linear with the frequency like in the DMI-type effect \cite{Zhang-2014}. The assessment of the interplay of these two effects is therefore crucial to disentangle their relative strength and to arrive to a clear understanding of the electric field effects on the magnetization dynamics.

In this paper we give a contribution to the topic by the investigation of the electric field effect in the context of micromagnetics. Our aim is to introduce the two mentioned electric field effects within the continuous description of the magnetization vector in which all the dynamic effects, as for example the spin waves, the domain wall motion and so on, can be properly described. Here we consider ferromagnetic insulators so that the electric field can be applied by the help of appropriate electrodes. As already mentioned, the effect of the electric field on the magnetization vector of the ferromagnet is twofold. 

The first effect \cite{Cao-1997, Basso-2020}, sometimes called the magneto electric effect or Aharonov-Casher effect, is of relativistic origin and is caused by the energy associated to the transport of a magnetic moment in an electric field. Within the micromagnetic theory this energy term has to be written as a function of the components of the magnetic moment current density tensor $\mathbf{j}_{\mathbf{M}}$. As a result the Aharonov-Casher effect on the spinwaves gives an additional phase proportional to the electric field $E$, i.e. the dispersion relation results to be shifted of the quantity $(\gamma_e/c^2)E$ where $\gamma_e$ is the gyromagnetic ratio and $c$ is the speed of light \cite{Basso-2020}.

The second effect, sometimes called the electric field induced DMI, is associated to the energy terms emerging from the broken inversion symmetry caused by the electric field \cite{Mills-2008}. Within micromagnetics, the appropriate approach to this second effect is a gauge theory applied to the exchange energy term of ferromagnetism \cite{Hill-2021, DiPietro-2022}. The result of the gauge approach is to provide two additional terms to the micromagnetic energy: i) a DMI type energy term, with coefficient $D$ plus ii) an hard axis anisotropy along the direction of the electric field with coefficient $D^2$ \cite{Perna-2022}.

\begin{figure}[htb]
\centering
\includegraphics[width=7cm]{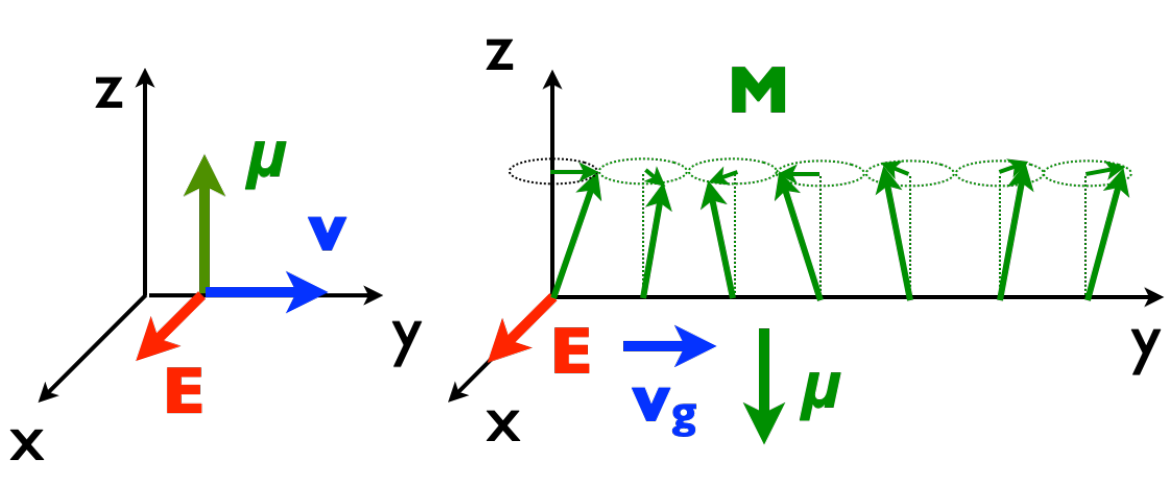}
\caption{Electric field effect on a particle with magnetic moment $\boldsymbol{\mu}$ and velocity $\mathbf{v}$ (left) and on a spin waves with main magnetization vector along $z$, transported magnetic moment in the opposite direction and group velocity $v_g$ along $y$ (right). The component of the electric field giving an effect is the one along $x$. } \label{FIG:moment}
\end{figure}

Both effects modify the dispersion relation of the spin waves, albeit in qualitatively different ways. Even if the spin wave dispersion relations with DMI are well known \cite{Moon-2013}, in order to describe the spin waves in ferromagnetic insulators under electric field we have to also consider the Aharonov-Casher shift and the additional DMI anisotropy obtained from the gauge theory. In this paper we extend the existing formalism with these two additional terms in the case of magnetostatic Damon-Eshbach surface waves \cite{Damon-1960, Eshbach-1960, Gurevich-1996, Stancil-2009} in which one has a thin film with thickness $d$ along $x$, main applied field and main magnetization direction along $z$ and propagation direction of the spin waves along $y$ (see Fig.\ref{FIG:moment}). We additionally consider the role played by the metallic electrodes, needed in the experiments, as boundary conditions for the magnetostatic field. The results are compared with the experimental data of the magnetic field and electric field dependence of the phase acquired by magnetostatic spin waves on ittrium iron garnet Y$_{\rm 3}$Fe$_{\rm 5}$O$_{\rm 12}$ (YIG) thin films from Ref.\cite{Zhang-2014} in which the phase acquired by the spin wave has two components: one is frequency independent as a consequence of the Aharonov-Casher effect and the other is linear with the frequency because of the DMI-type effects.

The main result of our study is that it is not possible to account for both effects by the same value of the applied electric field because the Aharonov-Casher effect would be slightly overestimated and the DMI-like contribution greatly underestimated. Therefore we introduce two dimensionless parameters $\alpha$ and $\beta$ given by the ratio of the electric field amplitude used to fit the data and the applied electric field $E_x$. The physical reason for this assumption is that the two effects are both sensitive to local electric fields that can possibly be very different in amplitude especially in a complex crystal like a ferrimagnetic oxide as YIG. Indeed the Aharonov-Casher effect should be associated to the average local electric field, $\alpha {E}_{x}$, felt by the magnetic moments at their atomic lattice sites. The result of the fit is $\alpha \simeq 1.2 \cdot 10^{-2}$ i.e. corresponding to a shielding of the applied electric field. The DMI type effect is associated to the ferromagnetic exchange interaction, therefore it will be proportional to the local average electric field in the bonding region between magnetic sites \cite{DiPietro-2022}. The corresponding field, $\beta {E}_{x}$, will also be proportional to the applied electric field ${E}_x$, however we may expect that the crystal symmetry breaking, induced by the applied electric field, brings about the contributions of the strong crystal field of the neighboring ions and amplifies the effect. The result for YIG is $\beta \simeq 10^{4}$. In the following sections we first introduce and discuss the Aharonov-Casher and the DMI effects in the context of micromagnetics and then discuss how their effects can be disentangled in the comparison with the experimental data on YIG.

\section{Electric field effect on ferromagnets}

We introduce the effects of an electric field within micromagnetics, the continuous description of the magnetization vector in which all the dynamic effects, as for example the spin waves, the domain wall motion and so on, can be properly described. We consider ferromagnetic insulators so that the electric field can be applied by the help of appropriate electrodes. As already mentioned, the effect of the electric field on the magnetization vector of the ferromagnet is twofold: i) the magneto electric effect or Aharonov-Casher effect \cite{Cao-1997, Basso-2020} and ii) the electric field induced DMI \cite{Mills-2008}.

\subsection{Aharonov-Casher effect}
\label{SECT:ME}

The Aharonov-Casher effect consists in the acquisition of an additional phase when a particle with a magnetic moment $\boldsymbol{\mu}$ is in motion with velocity $\mathbf{v}$ in presence of an electric field $\mathbf{E}$ \cite{Aharonov-1984}. This effect is dual to the Ahronov-Bohm effect for charged particles. The Aharonov-Casher effect is due to the fact that the canonical momentum of the particle is the sum of the kinetic momentum and of the electromagnetic momentum given by $- ({1}/{c^2}) \mathbf{E} \times \boldsymbol{\mu}$ (see Fig.\ref{FIG:moment} left). The consequence of the presence of the electromagnetic momentum is the acquisition of an additional phase when the particle traverses a region with an electric field. 

In a continuous description of ferromagnetism one finds an equivalent effect on magnetization waves or spin waves. Since spin waves have a finite group velocity and transport a net magnetic moment, then, when they propagate in an electric field they will also acquire an additional phase. To see this effect starting from the classical micromagnetic framework one has to follow an approach similar to the Aharonov-Casher case as it was done in Ref.\cite{Basso-2020}. The main steps are the following. The energy term associated to a magnetic moment current in an electric field is written as $ - (1/c^2) \epsilon_{ijk} E_k j_{{\rm M},ij}$, where $j_{{\rm M},ij}$ are the components of the magnetic moment current tensor (the first index $i$ is the direction of the current and the second index $j$ is the direction of the magnetic moment) and $\epsilon_{ijk}$ is the Levi-Civita tensor. The full Lagrangian of the system is obtained by adding this term to the Lagrangian of micromagnetism describing magnetic precession. From the complete Lagrangian it is not straightforward to write down the equation of motion because the expression of the magnetic moment current density is not explicit. However by limiting to the case of linear spin waves, one can show that the Lagrangian with the electric field can be expressed as the classical one provided one redefines the derivative operator \cite{Basso-2020}. For the spin waves dispersion relation, this corresponds to redefine the wavenumber. We choose the main magnetization vector along $z$ (with a magnetic field along $z$, $H_z$), the propagation direction along $y$ and the electric field along $x$ (see Fig.\ref{FIG:moment} right) and we get that any dispersion relation will be shifted in the wavenumber as  
 
\beq
q_y \rightarrow q_{y}+ (\gamma_e/c^2)E_x
\eeq

\noindent The shift applies to all kind of spin waves including magnetostatic waves. 

\begin{figure}[htb]
\centering
\includegraphics[width=5cm]{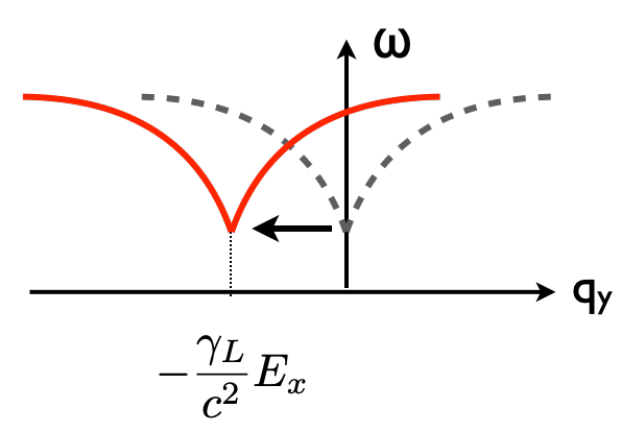}
\caption{Aharonov-Casher effect on spin waves. The picture shows the sketch of the dispersion relation for magnetostatic surface waves without (dashed gray line) and with (full red line) electric field (Eq.(\ref{EQ:shift})) } \label{FIG:shift}
\end{figure}

We consider here surface magnetostatic waves \cite{Gurevich-1996, Stancil-2009} with dispersion relation

\beq
\frac{\omega}{\mu_0\gamma_eM_s}= \sqrt{ \frac{H_0^2}{M_s^2} + \frac{1}{4} \left[ 1- \exp(- 2 d |q_{y}|)\right]}
\eeq

\noindent where $M_s$ is the saturation magnetization,

\beq
H_0 = \sqrt{H_z(H_z + M_s)}
\eeq

\noindent and $d$ is the thickness. Fig. \ref{FIG:shift} shows the sketch of the dispersion relation with the electric field shift. With small wavenumbers, the dispersion relation is approximated as a linear relation and by including the electric field shift we get

\beq
\frac{\omega}{\mu_0\gamma_eM_s} \simeq \frac{H_0}{M_s} + \frac{ M_s d}{4 H_0}  |q_{y}+ (\gamma_e/c^2)E_{x}|
\label{EQ:shift}
\eeq

\noindent An electric field $E_x = 10^{6}$ V/m with the constant $\gamma_e/c^2 \simeq 1.95 \cdot 10^{-6}$ V$^{-1}$, yields a wavenumber shift of about $1.95$ m$^{-1}$, corresponding to a small but measurable effect. 

\subsection{Electric field induced DMI}
\label{SECT:DMI}

The DMI is a chiral exchange interaction arising in presence of the breaking of inversion symmetry because of the crystal structure, the presence of adjacent non magnetic layers or the presence of an applied electric field \cite{Mills-2008}. 
To introduce the DMI for a ferromagnet within the micromagnetic approach we use a gauge field theory approach \cite{Tatara-2019} which has the advantage to fully exploit the symmetry properties of the system. To formulate the gauge theory of DMI we follow the approach of Refs.\cite{Hill-2021, DiPietro-2022}. The method consists in promoting the global symmetries of the system to local ones. The procedure is as follow. One has first to look for the global transformations which leave the Lagrangian invariant. For ferromagnets, these transformations are the rotations of the magnetization vector, namely the SO(3) group. Second, one has to introduce the corresponding local transformations and ensure the invariance under them (called gauge invariance) by introducing a covariant derivative dependent on the gauge field. By substituting the normal derivatives with the covariant ones the Lagrangian becomes invariant under local transformations but the corresponding gauge field is included in the physical description. In the case of ferromagnetism described by a continuous magnetization vector the covariant derivative $\mathcal{D}_i$ is defined as 

\beq
\mathcal{D}_i m_k= \partial_i m_k + \epsilon_{ljk} m_j D_{il} 
\eeq

\noindent where $\partial_i$ is the usual derivative, $m_k$ is the normalized magnetization and $D_{il}$ is the gauge field tensor. It we take only the antisymmetric part $D_{ij} = \epsilon_{ijl} D_{l}$ of the gauge field tensor and choose it along $x$, $D_x$, we find that the exchange energy of micromagnetism will be proportional to 

\begin{multline}
(\mathcal{D} \mathbf{m})^2= (\nabla \mathbf{m})^2 - 2 D_x ( m_x \partial_y m_y  - m_y \partial_y m_x) \\ + 2 D_x (m_z \partial_z m_x - m_x \partial_z m_z)+ D_x^2 m_x^2
\end{multline}

\noindent where, at the right hand side we get: i) the usual exchange, ii) the DMI-type energy term and iii) a DMI induced hard $x$ axis anisotropy term \cite{Perna-2022}. The gauge field $D_{x}$, measured in the same unit as the wavenumber, m$^{-1}$, quantifies the strength of the DMI effect. 

\begin{figure}[htb]
\centering
\includegraphics[width=5cm]{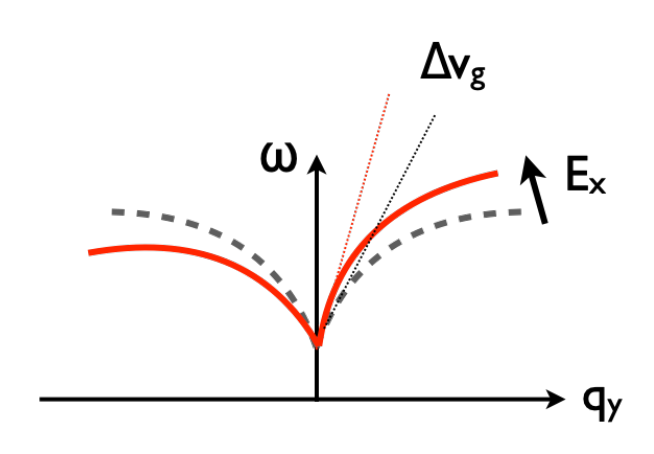}
\caption{Effect of the electric field induced DMI. The picture shows the sketch of the dispersion relation for magnetostatic surface waves without (dashed gray line) and with (full red line) electric field (Eq.(\ref{EQ:slope})) } \label{FIG:slope}
\end{figure}

Again we apply the previously derived effect to spin waves and we limit to consider surface magnetostatic waves. The dispersion relation with DMI and $x$ axis anisotropy, in the approximation of small $|q_{y}|$, is 

\beq
\frac{\omega}{\mu_0\gamma_eM_s} \simeq \frac{H_0}{M_s} + \frac{M_s-H_{AN}}{4 H_0}|q_{y}| d + 2l_{EX}^2D_x q_y
\label{EQ:slope}
\eeq

\noindent where we find at the right hand side a term linear in the wave vector $q_y$ as obtained in Ref.\cite{Moon-2013}, but we also get a DMI induced anisotropy term along $x$, $H_{AN} = - M_s (l_{EX}D_x)^2$ where $l_{EX} = [2A/(\mu_0 M_s^2)]^{1/2}$ is exchange length and $A$ is the exchange stiffness. The field $H_0$ is now $H_0 = \sqrt{H_z(H_z + M_s-H_{AN})}$. The main effect of the DMI-type energy terms is to change the group velocity of the spinwaves by $\Delta v_g = \mu_0\gamma_eM_s 2l_{EX}^2D_x$. Fig. \ref{FIG:slope} shows the sketch of the dispersion relation with the change of the slope due to the DMI.

From crystal symmetry considerations we expect that, for a centrosymmetric crystal like YIG, the DMI effect is absent without an applied electric field \cite{DiPietro-2022}, therefore we expect $D_x$ to be proportional to $E_{x}$. However if we would merely use the conversion factor between electric field and wavenumber found in the previous section giving $D_x = (\gamma_e/c^2)E_{x}$, we would greatly underestimate the DMI effect. To reconcile this fact we have to recall that the gauge effect just described acts at the level of exchange interaction and therefore the relevant electric field is the local average electric field felt in the bonding region. An estimate of the value of $D_x$ is therefore quite difficult to give and one should rely on first principle calculations or on the comparison with experimental data. We follow here this second approach and set $D_x = (\gamma_e/c^2) \beta E_x$, where $\beta$ is a coefficient to be determined. As a very rough order of magnitude estimate if we take $(\gamma_e/c^2) \beta = 1$ V$^{-1}$, we get, in YIG, with $\mu_0M_s \simeq 0.18$ T, $A \simeq 0.4 \cdot 10^{-11}$ J/m and $l_{EX} \simeq 18$ nm, a variation of the group velocity of $\Delta v_g \sim 20$ m/s in an applied electric field of $E_x = 10^{6}$ Vm$^{-1}$.

\section{Disentangling the electric field effects}
\label{SECT:DIS}

Having introduced the two distinct effects of an external electric field on the dispersion relation of the magnetostatic spin waves, Eqs.(\ref{EQ:shift}) and \eqref{EQ:slope}, we have now to combine them in order to have a comprehensive model to fit the experimental data. The main issue to address at this point is what we have introduced the effects within the continuous theory of micromagnetism, but we are going to apply them to the effect measured on a magnetic crystal. We have therefore also to discuss how to account for the effective electric field microscopically acting on the magnetic spins.

\subsection{Electric field effect in YIG crystals}


YIG has composition Y$_3$Fe$_{5}$O$_{12}$ and has a cubic crystal structure. The cations, Y$^{3+}$ and Fe$^{3+}$, are  surrounded by oxygen O$^{2-}$ ions forming the corners of geometric holes. Y$^{3+}$ ions are found in 8-cornered polyhedron holes and do not contribute to magnetism. Three out of the five Fe$^{3+}$ magnetic ions are in tetrahedral holes while the remaining two are in octahedral holes. Both Fe$^{3+}$ types contribute to magnetism with spin 5/2 each. Each O$^{2-}$ ion has two Y$^{3+}$ neighbor and two Fe$^{3+}$ ions, one in tetrahedral and one in octahedral holes. The two Fe$^{3+}$ sets form two magnetic sublattices with antiparallel magnetization and their main interaction is the superexchage mediated by the bonding oxygen O$^{2-}$ ion \cite{Serga-2010}. The effect of an applied electric field on such crystal structure can be rather complex. In particular it can produce an average local electric field upon the magnetic iron sites which is different from the applied one. Therefore the Aharonov-Casher effect must be described by an effective electric field $\alpha E_x$ where $\alpha$ is a dimensionless coefficient introduced to take into account the role of the crystal structure. For the electric field induced DMI effect we have already discussed that it is the result of the breaking of the inversion symmetry caused by the applied electric field and that the effect take place at the level of the ferromagnetic exchange interaction. The amplitude of the effect is far form being related the intensity of the applied field only and, in the case of YIG, the relevant electric field that contributes to the gauge field is at the level of the oxygen anion mediating the superexchange between the two magnetic Fe$^{3+}$ sublattices. The applied electric field may either slightly distort the crystal structure or modify the electronic wavefunctions in a complicated way. To account phenomenologically all these effects we introduce the DMI coefficient as $D_x = (\gamma_e/c^2) \beta E_x$ where $\beta$ is a dimensionless coefficient to be determined by the comparison with experiments.

\begin{figure}[htb]
\centering
\includegraphics[width=5cm]{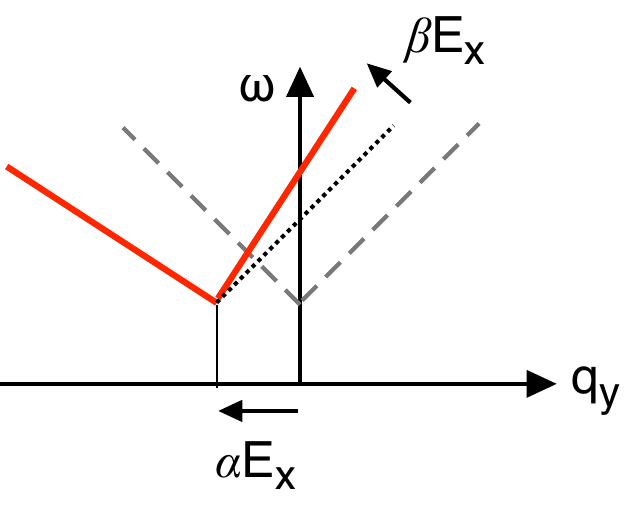}
\caption{Joint effect of the Aharonov-Casher phase and the electric field induced DMI. The picture shows the sketch of the dispersion relation for magnetostatic surface waves (in the linear approximation) without (dashed gray line) and with (full red line) electric field (Eq.(\ref{EQ:complete_model})). $\alpha$ and $\beta$ are two dimensionless pheomenolocical parameters introduced to account for the crystal symmetry breaking caused by the applied electric field.} \label{FIG:both}
\end{figure}

Combining equations \eqref{EQ:shift} and \eqref{EQ:slope}, the total electric field effect on the dispersion relation is given by

\begin{multline}
\frac{\omega}{\mu_0\gamma_eM_s} \simeq \frac{H_0}{M_s} + \frac{M_s-H_{AN}}{4 H_0}|q_{y}+ (\gamma_e/c^2)\alpha E_x| d \\ + 2l_{EX}^2(\gamma_e/c^2) \beta E_x (q_{y}+ (\gamma_e/c^2) \alpha E_x)
\label{EQ:complete_model}
\end{multline}

\noindent and schematically shown in Fig.\ref{FIG:both}. This dispersion relation shows that the phase acquired by the spin wave has two components: one is frequency independent because of the Aharonov-Casher effect and the other is approximately linear with the frequency because of the DMI-type effect. 

\subsection{Spin waves in YIG}

\begin{figure}[htb]
\centering
\includegraphics[width=9cm]{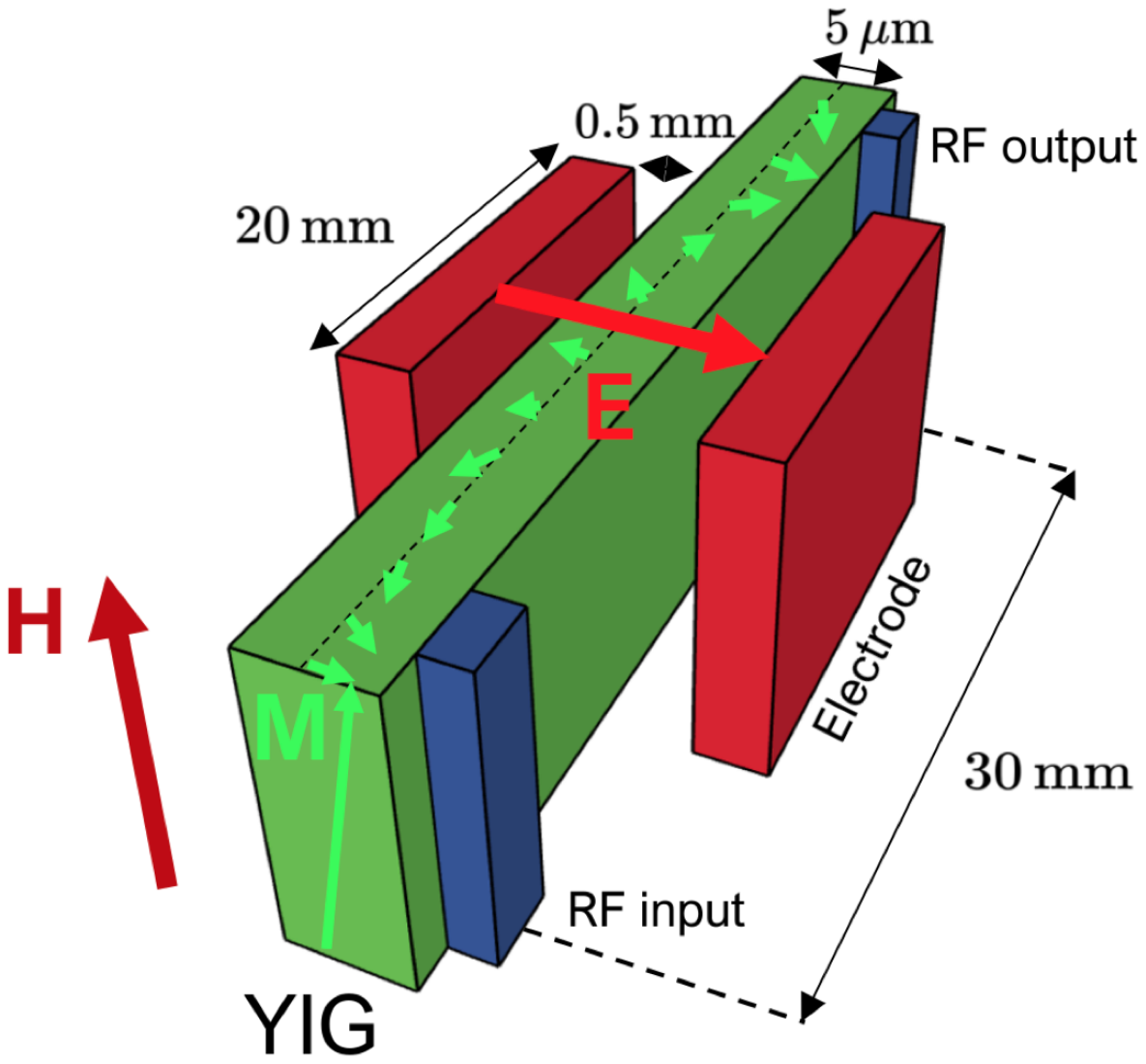}
\caption{Sketch of the YIG waveguide of Ref.\cite{Zhang-2014}. The YIG single crystal (green) has thickness of 5 $\mu$m. Spinwaves are measured by two RF antennae (blue) at a distance of 30 mm. The electric field is applied by means of electrodes (red) of length 20 mm. The front electrode is almost in contact with the YIG (the resulting gap from the fit is $34\, \mu$m) the back electrode is in contact with the GGG substrate (not shown) and the resulting gap is $0.5$ mm.} \label{FIG:exp}
\end{figure}

\begin{figure}[htb]
\centering
\includegraphics[width=8cm]{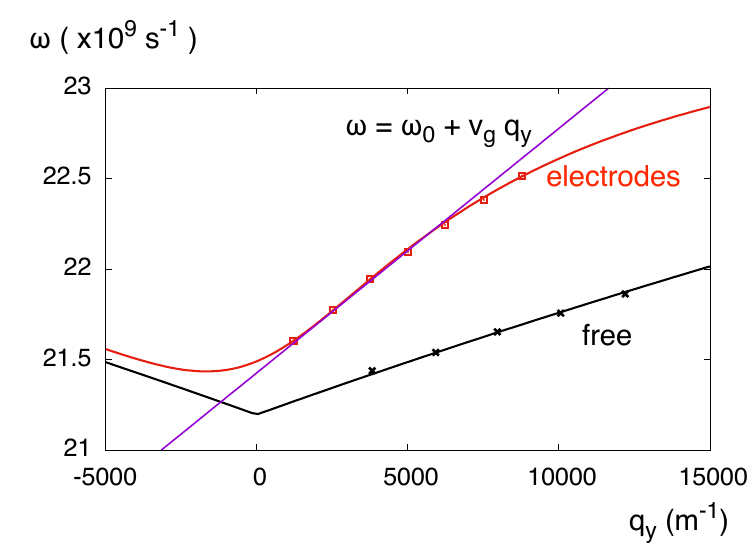}
\caption{Dispersion relation of surface waves for a YIG film of thickness $d=5\,\mu$m. Points: experimental data from Ref.\cite{Zhang-2014}. Black crosses: free propagation (the original data is shifted by the offset $\Delta q_y = 3800$ m$^{-1}$); red squares: metallic electrodes. Lines: theory from the solution of Eq.(\ref{EQ:eq}). Black line: free propagation; red line metallic electrodes. The curves are obtained with the following parameters: $\mu_0M_s = 0.181$ T, $\mu_0 H_z = 60.1$ mT, $g_- = 0.5$ mm and $g_+ = 34\, \mu$m. The purple line is a linear fit of the solution with metallic electrodes.} \label{FIG:dispersion}
\end{figure}

The experiments of Ref.\cite{Zhang-2014} were performed on a YIG single crystal of thickness of 5 $\mu$m grown on GGG. The dispersion relation was measured by placing two RF antennae at a distance of 30 mm and by measuring the phase $\Delta \varphi$ acquired by the spinwave from the transmitting to the receiving antenna. The electric field is applied by adding electrodes in the central part of the YIG wave guide as shown in Fig.\ref{FIG:exp}. The presence of the metallic electrodes changes the dispersion relation as shown in Fig.\ref{FIG:dispersion} where the points are the measured data from Ref.\cite{Zhang-2014}. The theoretical fit of the data of Fig.\ref{FIG:dispersion} is obtained by using the dispersion relation with and without metallic electrodes shown in Appendix \ref{APP:BOUND}. In the free case (without electrodes) the fit was obtained by adding an offset to the data of $\Delta q_y = 3800$ m$^{-1}$ due to the fact that the experimental method does not permit to address the starting point of the dispersion relation. 

When the electric field was applied, the acquired phase $\Delta \varphi$ displays a small change linear with $E_x$. As $\Delta \varphi = - L_y\Delta q_y$ where $L_y = 20$ mm is the length of the electrodes, we can derive the electric field induced change in the wave number $\partial q_y/\partial E_x$ as a function of the exciting frequency $\omega$. The experimental data shows that $\partial q_y/\partial E_x$ has two contribution: one frequency independent at $\omega_0$ and one frequency dependent proportional to $(\omega-\omega_0)$ where $\omega_0$ is the resonance frequency of the uniform mode. The first one is the consequence of the Aharonov-Casher effect while the second of the electric field induced DMI. Our point here is to understand their relative amplitudes.
 
As a first baseline case, we discuss the limit of the model $\alpha = \beta = 1$, which implies that all the effects on the dispersion relation operate on the same scale and are a consequence of the direct applied electric field effect. We immediately notice how this assumption leads to nonphysical results as it would end up overestimating the influence of the AC phase acquisition and greatly underestimating the DMI-like contribution. The inclusion of the $\alpha$ and $\beta$ coefficients on the other hand, allows to disentangle the two effects and treat them on the respective relevant scales. The experimental value of the frequency independent change is, at $E_x = 10^{6}\,\, \mbox{V/m}$, $\Delta q_y(\omega_0) = 2.5 \cdot 10^{-2}$ m$^{-1}$ while the measured frequency dependent change is

\begin{equation}
\frac{\partial \Delta q_y}{\partial \omega} = \Delta \left( \frac{1}{v_g} \right)= 3.2 \cdot 10^{-11} \,\, \mbox{s/m}
\end{equation}

\noindent We obtain therefore $\alpha = 1.2 \cdot 10^{-2}$ and $\beta = 10^4$, highlighting the vastly different scales at which the effective electric fields operate. We also recognize how, with these values, the DMI induced anisotropy component $(l_{EX}D_x)^2/M_s$ is of order $10^{-7}$  and can therefore be disregarded in the present case.


\section{Discussion and conclusions}

The main result of the inclusion of the electric field in the dispersion relation of the magnetostatic spin waves is given by Eq.(\ref{EQ:complete_model}). The comparison with experimental data on YIG of Ref.\cite{Zhang-2014} however highlights that we cannot employ the value of the applied electric field to get the experimentally obtained dispersion relations without a slight overestimation of the Aharonov-Casher contribution and a great underestimation of the DMI-like one. Therefore by introducing the $\alpha,\beta$ parameters as in eq.\eqref{EQ:complete_model} we get an estimate of the effective electric field acting locally on the magnetic ions. With an applied electric field of $E_x = 10^{6}$ V/m we obtain that the electric field of Eq.(\ref{EQ:shift}), interpreted as an effective electric field perceived by the traveling spin wave, has order of magnitude $\alpha E_x \sim 10^4$ V/m while the electric field necessary to reproduce the required tilt as a consequence of the DMI effect on magnetic exchange, Eq.(\ref{EQ:slope}), must have order of magnitude of $\beta E_x \sim 10^{10}$ V/m. To address this difference, we recall how according to the discussion in sections \ref{SECT:ME} and \ref{SECT:DMI}, the very nature of the electric field appearing in the dispersion relation needs careful consideration. In the case of the Aharonov-Casher, we are dealing with something closer to the applied electric field, perhaps mediated by some partial screening of the ions of the crystal. In the case of the DMI, the relevant local electric field may have strong contributions from the inversion symmetry braking of the local crystal electric field. Of course, in the absence of an applied electric field, the centrosymmetry of YIG forbids any form of DMI like exchange.

In conclusion, we have shown that through a micromagnetic theoretical analysis of the electric field effect on the dispersion relation of magnetostatic surface waves, two main contributions emerge: a purely macroscopic relativistic one and a contribution that operates on the level of quantum mechanical magnetic exchange. We speculate that since the physical origin of these two is drastically different, the nature of the electric field itself must be carefully considered when analyzing experimental data: in the case of the Aharonov-Casher contribution, an effective electric field of an order of magnitude similar to the applied electric field can be considered. For the contribution operating on the level of magnetic exchange, a much stronger electric field must be considered whose origin lies in the crystal symmetry breaking due to the applied electric that brings about the contributions of the crystal field of the neighboring ions.

\appendix
\section{Magnetostatic waves with metallic boundaries}
\label{APP:BOUND}

The application of an electric field along $x$ requires the presence of metallic electrodes that also act as metallic boundary conditions for the magnetostatic field. The effect of metallic boundaries on surface waves has been studied by Bongianni \cite{Bongianni-1972}, Yukawa \cite{Yukawa-1977}, and O'Keeffe and Patterson \cite{OKeeffe-1978}. If we also include the presence of an anisotropy along $x$ we find the local inverse susceptibility matrix given by 

\beq
\bar{\chi}^{-1}=
\frac{1}{M_s}\left[ \begin{array}{cc}
H_z - H_{AN} & i \omega/(\mu_0\gamma_e) \\
- i \omega/(\mu_0\gamma_e) & H_z
\end{array} \right]
\eeq

\noindent By inverting the matrix we write the susceptibility as

\beq
\bar{\chi}=
\left[ \begin{array}{cc}
\chi_x & - i \kappa \\
i \kappa & \chi_y
\end{array} \right]
\eeq

\noindent and we find the dispersion relation by solving the equation

\beq
[\mu^2 - \kappa^2 + t_+t_- - \nu\kappa (t_+-t_-) ] t_d + \mu (t_++t_-)=0
\label{EQ:eq}
\eeq

\noindent where $\mu=\sqrt{(1+\chi_x)(1+\chi_y)}$, $\nu = \pm1$ is the traveling direction, $t_d=\tanh(d|q_{y}|)$, $t_{\pm}=\tanh\left(g_{\pm}|q_{y}| \right)$ and $g_{\pm}$ are the air gaps between the ferromagnet and the electrodes. Fig.\ref{FIG:dispersion} shows as an example the dispersion relation of YIG with $\mu_0M_s = 0.181$ T, $d=5\,\mu$m, $\mu_0 H_z = 60.1$ mT. The electrodes are at $g_- = 0.5$ mm and $g_+ = 34\, \mu$m. The chosen values correspond to the experiments of Ref.\cite{Zhang-2014}. We can see that the presence of the electrodes strongly increases the group velocity $v_g$ for waves traveling in the positive direction ($\nu=1$) going from $v_g \simeq 0.52 \cdot 10^5$ m/s for the free case to $v_g \simeq 1.35 \cdot 10^5$ m/s for the metallic electrodes case.

 \section*{Acknowledgement} The authors gratefully acknowledge the financial support from the European Union's Horizon 2020 research and innovation programme under the Marie Sklodowska-Curie grant agreement No. 860060 "Magnetism and the effect of Electric Field" (MagnEFi).

\bibliographystyle{model1-num-names}
\bibliography{00BibQ}
\end{document}